\documentclass[12pt,preprint]{aastex}

\newcommand{\rcb}{$r_{\rm CB}$}
\newcommand{\rcr}{$r_{\rm CR}$}
\newcommand{\rkep}{$r_{\rm Kep}$}

\shorttitle{Infalling-Rotating Envelope around IRAS 04365+2535}
\shortauthors{Sakai et al.}

\usepackage{lscape}
\begin{document}


\title{Subarcsecond Analysis of Infalling-Rotating Envelope around the Class I Protostar IRAS 04365+2535}


\author{Nami Sakai\altaffilmark{1}, Yoko Oya\altaffilmark{2}, Ana L\'{o}pez-Sepulcre\altaffilmark{2}, Yoshimasa Watanabe\altaffilmark{2}, Takeshi Sakai\altaffilmark{3}, Tomoya Hirota\altaffilmark{4}, Yuri Aikawa\altaffilmark{5}, Cecilia Ceccarelli\altaffilmark{6}, Bertrand Lefloch\altaffilmark{6}, Emmanuel Caux\altaffilmark{7}, Charlotte Vastel\altaffilmark{7}, Claudine Kahane\altaffilmark{6}, and Satoshi Yamamoto\altaffilmark{2}}
\altaffiltext{1}{The Institute of Physical and Chemical Research (RIKEN), 2-1, Hirosawa, Wako-shi, Saitama 351-0198, Japan}
\altaffiltext{2}{Department of Physics, The University of Tokyo, Bunkyo-ku, Tokyo 113-0033, Japan}
\altaffiltext{3}{Department of Communication Engineering and Informatics, Graduate School of Informatics and Engineering, The University of Electro-Communications, Chofugaoka, Chofu, Tokyo, 182-8585, Japan}
\altaffiltext{4}{National Astronomical Observatory of Japan, Osawa, Mitaka, Tokyo 181-8588, Japan}
\altaffiltext{5}{Center for Computational Science, University of Tsukuba, Tsukuba, Ibaraki 305-8577, Japan }
\altaffiltext{6}{Universite de Grenoble Alpes, IPAG, F-38000 Grenoble, France; CNRS, IPAG, F-38000 Grenoble, France}
\altaffiltext{7}{Universite de Toulouse, UPS-OMP, IRAP, Toulouse, France; CNRS, IRAP, 9 Av. Colonel Roche, BP 44346, F-31028 Toulouse Cedex 4, France}

\begin{abstract}
Sub-arcsecond images of the rotational line emission of CS and SO have been obtained toward the Class I protostar IRAS 04365$+$2535 in TMC-1A with ALMA.  A compact component around the protostar is clearly detected in the CS and SO emission.  The velocity structure of the compact component of CS reveals infalling-rotating motion conserving the angular momentum.  It is well explained by a ballistic model of an infalling-rotating envelope with the radius of the centrifugal barrier (a half of the centrifugal radius) of 50 AU, although the distribution of the infalling gas is asymmetric around the protostar.  The distribution of SO is mostly concentrated around the radius of the centrifugal barrier of the simple model.  Thus a drastic change in chemical composition of the gas infalling onto the protostar is found to occur at a 50 AU scale probably due to accretion shocks, demonstrating that the infalling material is significantly processed before being delivered into the disk.
\end{abstract}


\clearpage

\section{Introduction}
In the formation of low-mass stars, a rotationally-supported disk (Keplerian disk) is formed around a newly born protostar from an infalling envelope 
due to a natural consequence of angular momentum conservation. 
It evolves to a protoplanetary disk and eventually to a planetary system. 
A detailed understanding of the process is thus a central issue for star-formation studies. 
Tracing chemical evolution during this process is also important in exploring the origin of the Solar System. 
Although there is strong evidence for rotationally-supported disks toward Class I sources and a few Class 0 sources 
(e.g. Hogerheijde 2001; Brinch et al. 2007; Takakuwa et al. 2012; Tobin, J. et al. 2012; Murillo et al. 2013), 
it is still controversial when and how the disk structure is formed (e.g. Li, Krasnopolsky, and Shang 2011; Machida, Inutsuka, and Matsumoto 2014; Tsukamoto et al. 2015).

Recently, Sakai et al. (2014a; b) reported the sub-arcsecond observation of various molecular lines toward the Class 0 protostar IRAS 04368$+$2557 in L1527 with ALMA, 
and discovered a drastic chemical change at the centrifugal barrier of the infalling-rotating envelope, which is a perihelion radius of a infalling particle (a half of the centrifugal radius). 
CCH, c-C$_3$H$_2$, and CS preferentially exist outward of the centrifugal barrier of 100 AU, 
while SO only exists in a ring-like structure around the centrifugal barrier due to liberation of the SO ice 
by the accretion shock (Aota, Inoue, and Aikawa 2015). 
Inward of the centrifugal barrier, CCH, c-C$_3$H$_2$, and CS will be 
frozen-out onto dust grains in a very short time scale ($\sim$100 yr:  See Aikawa et al. (2012) for its evaluation) in the mid-plane of the disk structure 
because of low temperature and high H$_2$ density (Sakai et al. 2014a). 
On the other hand, a weak emission whose velocity exeeds the rotation velocity at the centrifugal barrier 
is also observed for SO and H$_2$CO, and marginally for CS, in addition to the envelope and/or ring components, suggesting the existence of a Keplerian disk. 
Indeed, Ohashi et al. (2014) reported an indication of the Keplerian disk with a radius of 54 AU on the basis of the rotation velocity profile of the C$^{18}$O ($J$=2--1) emission observed with ALMA. 
In their C$^{18}$O observation, the centrifugal barrier is not identified, probably because CO can exist everywhere from the envelope to the disk. 
The above species are much more sensitive to the physical phenomena occurring around the protostar than CO, thus highlighting the existence of the centrifugal barrier. 

In the present study, we extend such a chemical approach to the low-mass Class I source 
IRAS 04365+2535 in TMC-1A, 
whose luminosity is 2.5 $L_\odot$ (Green et al. 2013). 
This source is subject to extensive studies exploring its disk structure. 
While the observations with relatively low spatial resolution ($>$ a few arcsecond) suggest the existence of a large ($\sim$several hundreds AU) disk with Keplerian rotation (Ohashi et al. 1997; Brown and Chandler 1999; Yen et al. 2013), recent sub-arcsecond resolution observations with PdBI suggest that the radius of the disk is at most 100 AU, on the basis of the power-law index ($-$0.5) of the rotation profile (Harsono et al. 2014).  The central mass is thus estimated to be 0.53 $M_\odot$.  Here, the inclination angle of the disk is assumed to be 35\degr (0$^\circ$ for edge-on), 
where the positive angle means that the upper side of the envelope faces us. 
Very recently, Aso et al. (2015) reported the existence of a Keplerian disk with a size of 100 AU on the basis of the rotation profile of the C$^{18}$O line observed with ALMA ($0\farcs9 \times 0\farcs9$). 

TMC-1A is located in the Heiles Cloud 2 ($d$=137 pc; Torres et al. 2007),  
where the carbon-chain-molecule rich starless core TMC-1 and the warm-carbon-chain-chemistry (WCCC) source L1527 (Sakai et al. 2008) are associated. 
Since high excitation lines of carbon-chain molecules such as CCH ($N$=3--2) and HC$_5$N ($J$=32--31) are detected in TMC-1A with the IRAM 30 m telescope (Sakai et al. in prep.), it seems likely that TMC-1A is an evolved (Class I) WCCC source. 
We already demonstrated that CS and SO are good tracers for the existence of a centrifugal barrier in the WCCC source L1527 (Sakai et al. 2014a; b). 
We then conducted the sub-arcsecond resolution observation of these molecules toward TMC-1A with ALMA 
in order to characterize physics and chemistry of the infalling-rotating envelope in a more evolved stage.\\

\section{Observations}
The observation of IRAS 04365$+$2535 in L1527 was carried out with ALMA (cycle 2) on 2014 June 15 (Band 6). 
The observed lines are CS ($J$=5--4; 244.9355565 GHz $E_{\rm u}$=35 K) and SO ($J_N$=$7_6$--$6_5$, 261.843721 GHz $E_{\rm u}$=48 K).  
Thirty-one antennas were used during the observation, 
and the maximum baseline length was 642 m. 
The field center is: ($\alpha_{2000}, \delta_{2000})$=$(04^{\rm h} 39^{\rm m} 35\fs2, 25^{\circ} 41^{\prime} 44\farcs35)$. 
The typical system temperature in Band 6 was 60--100 K. 
The backend correlator was tuned to be 59 MHz bandwidth with 122 kHz resolution (channel spacing is 61 kHz). 
The velocity resolution is 0.14 km s$^{-1}$. 
J0510$+$1800 was used for the phase calibration every 9 minutes. 
The bandpass calibration was carried out on J0510$+$1800. 
The absolute flux density scale was derived from J0423-013. 
The data calibration was performed in the antenna-based manner, and uncertainties are less than 10 \%. 
We did not conduct self-calibration for the data used in this paper. 
The continuum image was prepared by averaging line-free channels. 
The line images were obtained by CLEANing the dirty images after subtracting the continuum directly from the visibilities. 
Here, we used the Briggs weighting for the UV data with the robust parameter of 0.5. 
The primary beam (HPBW) is 23$^{\prime\prime}$. 
The total on-source time was 27 min. 
The synthesized beam size is $0\farcs65 \times 0\farcs37$ (PA=13\degr), 
$0\farcs69 \times 0\farcs40$ (PA=11\degr), and $0\farcs65 \times 0\farcs37$ (PA=12\degr), for the continuum, CS, and SO images, respectively. \\

\section{Results and Discussions}
\subsection{Continuum}
Figure 1a shows an image of the continuum emission at 260 GHz. 
The peak position derived by the two-dimensional Gaussian fit is $(\alpha_{2000}, \delta_{2000})$=$(04^{\rm h} 39^{\rm m} 35\fs2, 25^{\circ} 41^{\prime} 44\farcs19)$. 
The size of the distribution is $0\farcs75 \times 0\farcs52$ with a position angle of 19$^\circ$. 
Hence, the image is marginally resolved ($0\farcs40 \times 0\farcs32$ for the deconvolved size).  
The peak flux is determined to be 166.9 $\pm$0.9 mJy beam$^{-1}$ from the Gaussian fit, while the integrated flux is 268.2$\pm$2.3 mJy. 
The integrated flux is almost consistent with those reported by Harsono et al. (2014) (164 mJy at 220 GHz), and Aso et al. (2015) (182.1 mJy at 225 GHz), considering the difference of the frequency and a $\beta$ index of 1.46 (Chandler et al. 1998). \\ 

\subsection{CS}
Figure 1b shows the moment 0 map of CS ($J$=5--4). 
The CS distribution is extended toward the northeastern direction, 
and also shows clear concentration to the protostar with a size scale of 300 AU. 
The distribution is fairly asymmetric even in the centrally concentrated component, where the northeastern part is brighter. 
The extended component in the northeastern part has a typical velocity of 5.8 km s$^{-1}$. 
A systemic velocity of this source is 6.5 km s$^{-1}$, which is an average of the reported values (Brown \& Chandler 1999; Yen et al. 2013; Harsono et al. 2014), 
and hence, 
the 5.8 km s$^{-1}$ component traces the blue shifted gas associated with the protostar. 
The corresponding red-shifted component does not appear, probably linked to the asymmetric gas distribution. 

Figure 1b shows the moment 1 map superposed on contours of the moment 0 map. 
A clear velocity gradient can be seen, where the northeastern and southwestern parts are blue-shifted and red-shifted, respectively. 
This clearly confirms the rotation motion of the disk/envelope system around the protostar. 
Figure 1a shows the distributions of the most blue-shifted component and the most red-shifted component (3.5 km s$^{-1}$ $>|$$\delta$V$|>$ 2.5 km s$^{-1}$) on the continuum distribution, where $\delta V$ denotes the velocity shift from the systemic velocity. 
The peaks of these two components do not coincide with the peak of the continuum emission, and these three peaks are aligned on a single line. 
This situation is similar to that observed for the C$^{18}$O ($J$=2--1) emission by Aso et al. (2015). 
We determined the position angle (PA) of the disk/envelope system to be 70$^\circ$ using this line. 
It is consistent with that reported by Harsono et al. (2014) (-115$^\circ$, corresponding to 65$^\circ$ in our definition) and that reported by Aso et al. (2015) (73$^\circ$). 

The position-velocity (PV) diagrams along the envelope direction and along the line perpendicular to it through the protostar position (outflow direction) are shown in Figure 2a. 
In the PV diagram along the envelope, a clear spin-up feature is seen toward the protostar. 
The rotation motion traced by CS abruptly disappears around the radius of 50 AU from the protostar, while rotation is still evident to smaller radii in the C$^{18}$O emission (Aso et al. 2015). 
In addition, a counter velocity component caused by the infalling motion is seen in the northeastern part. 
Although the corresponding counter velocity component in the southwestern part is missing, this may be due to the asymmetrical distribution of the infalling gas around the protostar. 
More importantly, the velocity gradient is evident along the outflow direction. 
All these characteristic features seen in the CS emission cannot be explained by a Keplerian disk, but in contrast, strongly suggest an infalling-rotating envelope. 

Here, we analyzed the PV diagram with a simple ballistic model of an infalling-rotating envelope which is previously applied to L1527 (Sakai et al. 2014a, b; Oya et al. 2015) and IRAS15398-3359 (Oya et al. 2014). 
In the ballistic model, the infalling velocity takes its maximum at the centrifugal radius, 
\rcr\ (twice the radius of the centrifugal barrier, \rcb), where the infalling velocity equals to the rotation velocity.  
The rotation velocity increases by a factor of 2 from \rcr\ to \rcb. 
Toward the protostar position, only the infalling velocity can be measured, and we find that its maximum velocity at the redshifted side is just a half of the rotation velocity at 50 AU (Figure 2a).  
This means that the gas is certainly infalling beyond \rcr\ at least in the southwestern side.  
The radius (50 AU), where the CS emission tracing rotation motion abruptly disappears in the southwestern side, 
is thus interpreted as the radius of the centrifugal barrier \rcb\ in our model, 
because of the above factor-of-2 relation. 

We performed simulations of molecular emission using the simple ballistic model.  In the model,  the gas motion is approximated by the particle motion under conservation of kinetic energy and angular momentum. 
Under this assumption, the gas motion depends on the two model parameters, 
\rcb\ and the total mass of the protostar and the inner disk ($M$) (Oya et al. 2014). 
Since our main purpose is to account for the velocity structure, we simply assume an $r^{-1.5}$ density profile and constant molecular abundances. 
The lines are assumed to be optically thin, and no excitation effect is considered for simplicity. 
The only difference from the previous models is that we considered an increase in the envelope thickness toward the outside, as shown in a schematic view of Figure 1d. 
The extension angle of the envelope thickness is assumed to be 30$^\circ$, judging from the K-band (1.92-2.55 $\mu$m) image of this source (Ishii, Tamura, and Itoh 2004). 
  
We investigated a fairly large space for the three parameters, $r_{\rm CB}$, $M$, and the inclination angle of the envelope. 
We assumed an outer radius of the envelope of 350 AU, which is the size of the CS emitting region around the protostar (Figure 1b). 
The model image is convolved by the velocity width of 0.5 km s$^{-1}$ and the beam size of $0\farcs69 \times 0\farcs40$ (PA=11$^\circ$). 
Although we chose this velocity width by considering the turbulent motion expected in front of the centrifugal barrier, 
the simulation does not change significantly by use of the 0.2 km s$^{-1}$ line width.  
We found a set of parameters which can reasonably reproduce the observed PV diagrams. 
The acceptable range of $r_{\rm CB}$ is from 40 AU to 60 AU, and that of the mass is from 0.20 $M_\odot$ to 0.30 $M_\odot$. 
The inclination angle (i) is from 10\degr\ to 30\degr, 
which is consistent with those reported by Harsono et al. (2014) and Aso et al. (2015) 
($25\degr$ and $35\degr$, respectively, in our definition). 
The result of the simulation with the most probable parameters ($r_{\rm CB}$=50 AU, $M$=0.25 $M_\odot$, and $i$=20\degr) are shown in Figures 2a and 2b.  
The model well reproduces the characteristic features of the observed PV diagrams. 

On the basis of a change in the power-law index of the rotation curve, 
Harsono et al. (2014) and Aso et al. (2015) reported the radius of the Keplerian disk \rkep\ to be about 100 AU, 
which is larger than \rcb\ derived by our simple model. 
Since an infalling fluid parcel cannot simply cross the disk to reach \rcb\ in hydrodynamics, 
our ballistic model could be oversimplified. 
Detailed analysis of more realistic (and complex) gas motion in the infalling-rotating envelope and the transition zone to the disk is desirable (e.g. Stahler et al. 1994), 
but is out of the scope of the present study. 
Our observation and analysis clearly show, however, an infalling flow down to $\sim$50 AU. 
Since it is smaller than \rkep\ reported, it could be an accretion flow from the envelope onto the disk. 
In the following, we still refer to this radius as \rcb\ for simplicity. 

It is also interesting to compare the protostellar mass between our model and Aso et al. (2015). 
We derived the protostellar mass to be 0.25 $M_\odot$ from a simple analysis of an infalling-rotating envelope, 
while Aso et al. (2015) analyzed the Keplerian disk inside the envelope to derive the mass of 0.68 $M_\odot$ (inclination corrected). 
This disagreement is in fact consistent with the findings by Aso et al. (2015); in their model using the protostellar mass of 0.68 $M_\odot$, 
the infalling speed of the infalling-rotating envelope is evaluated to be higher than their C$^{18}$O ($J$=2--1) observations. 
Then, they introduce an effective parameter to reduce the infalling speed to explain the kinematics of the infalling-rotating envelope part. 
Hence, structure and physics of the transition zone from the infalling-rotating envelope to the Keplerian disk still remain to be explored
for thorough understandings of disk formation from the envelope.

\subsection{SO}
Figure 1c shows the moment 0 map of the SO ($J_N$=$7_6$--$6_5$) line. 
The critical density of this line calculated from the Einstein A coefficient and the collisional cross section (LAMDA; Sch$\rm \ddot{o}$ier et al. 2005; Lique et al. 2006) is $\sim 10^7$ cm$^{-3}$, which is similar to that of the CS ($J$=5--4) line. 
Nevertheless, the distribution is concentrated around the protostar, and is more compact than that of CS. 
The SO intensity is brighter in the southern part of the integrated intensity map in contrast to the CS case. 
Figures 2c and 2d show the PV diagrams along the envelope direction and the outflow direction. 
Along the envelope, the SO intensity is brighter in the southeastern part of the PV diagram than in the northeastern part. 
Since the CS emission is brighter in the northeastern part, the intensity of SO seems to be anticorrelated with that of CS. 
In the southwestern part, the SO emission 
is concentrated around the radius of the centrifugal barrier, as in the case of L1527. 
The SO molecule would be liberated into the gas phase from ice mantle by a weak accretion shock near the centrifugal barrier. 
The actual shock would be occurring in a certain range of the radius in front of the centrifugal barrier, 
possibly between \rcb\ and \rcr. 
This situation can be seen in the northeastern side, as described below. 
Figures 2c and 2d show the model of the infalling-rotating envelope with the same parameters mentioned above. 
Enhancement of SO near the centrifugal barrier can be recognized. 

On the other hand, the SO distribution is slightly extended outward of the centrifugal barrier in the northeastern part up to $\sim$100 AU. 
As indicated by the CS distribution, the envelope gas distribution is asymmetric around the protostar, 
and a large amount of the gas is stagnated in front of the centrifugal barrier in the northeastern part. 
As the result, the accretion shock would occur at a larger radius than the radius of the centrifugal barrier. 
Hence, the shock would be weakened resulting in a weak SO intensity there. 
Alternatively, a part of SO may originate from thermal evaporation by protostellar heating. 
According to the model by Brown and Chandler (1999), 
the `disk' temperature could be as high as $155 \pm 99$ K at 100 AU. 
This estimate has a large uncertainty, and the actual temperature would be somewhat lower according to the general model for the Class I sources (Whitney et al. 2003). 
Nevertheless, an inner part of the envelope might have higher temperature than the evaporation temperature of SO ($\sim$60 K). 

The abundance ratio of SO/CS at the centrifugal barrier is derived by using the same non-LTE method described in Sakai et al. (2014b). 
For this purpose, we used the intensities integrated from 7.9 to 9.3 km s$^{-1}$ and from 3.7 to 5.1 km s$^{-1}$ for the red-shifted and blue-shifted components, 
because the systemic velocity component would suffer from self-absorption. 
The SO/CS ratio is evaluated to be 4.6$-$8.8 and 1.9$-$3.2 for the red-shifted and blue-shifted components 
for the H$_2$ density range from $3 \times 10^6$ to $3 \times 10^8$ cm$^{-3}$ and the temperature range from 30 to 120 K. 
The ratio for the red-shifted component is comparable to that found in L1527 (4.6$-$5.9), while that for the blue-shifted component is slightly lower. 
The latter result would reflect the weaker accretion shock mentioned above.

Apart from the asymmetry of the envelope gas, the physical and chemical situation seen in this source is similar to that observed in L1527. 
Namely, CS is mostly distributed outward of the centrifugal barrier, while SO is enhanced around and in front of the centrifugal barrier. 
We also observed the CCH and SO$_2$ lines, 
whose distributions are similar to that of CS and SO, respectively (Sakai et al. in prep.). 
Thus, chemistry highlights the centrifugal barrier. 
Inward of the centrifugal barrier, 
these molecules would be depleted onto dust grains in the mid-plane of the disk forming region, although a part of them, 
particularly those near the surface of the disk, will stay in the gas phase.

\section{Concluding Remarks}
We analyzed the kinematic structure of the infalling-rotating envelope traced by the CS line by using the simple ballistic model. 
The PV diagrams are effectively explained by the simple model, 
whose centrifugal barrier is 50 AU. 
Although this source is in the Class I stage, the physical and chemical picture is quite similar to 
that observed in the Class 0 source L1527 except for a smaller amount of the infalling gas and its asymmetric distribution. 
A drastic chemical change is occurring around the radius of the centrifugal barrier (50 AU). 
This means that the envelope materials are significantly processed before being delivered into the disk component. 
The present study indicates that the detailed characterization of the envelope is important to study disk formation, 
and that chemistry is a powerful tool for this purpose. 

\acknowledgments

This paper makes use of the following ALMA data: ADS/JAO.ALMA\#2013.0.01102.S. ALMA is a partnership of ESO (representing its member states), NSF (USA) and NINS (Japan), together with NRC (Canada) and NSC and ASIAA (Taiwan), in cooperation with the Republic of Chile. The Joint ALMA Observatory is operated by ESO, AUI/NRAO and NAOJ. 
This study is supported by Grant-in-Aids from Ministry of Education, Culture, Sports, Science, and Technologies of Japan (21224002, 25400223, and 25108005). 
The authors acknowledge financial supports by JSPS and MAEE under the Japan-France integrated action programme (SAKURA: 25765VC).  \\





\begin{figure}
\epsscale{1.0}
\plotone{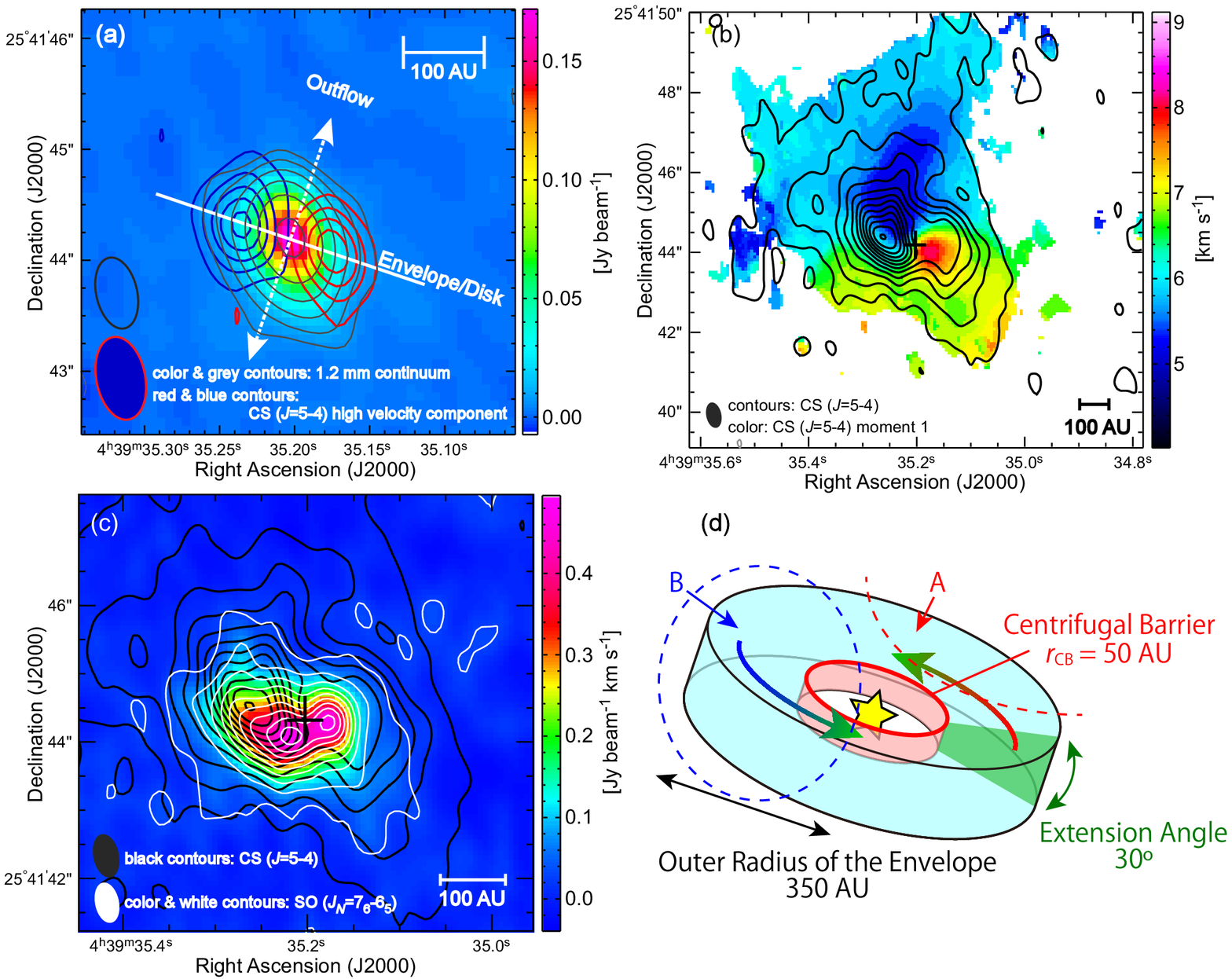}
\caption{(a) Continuum image (black contours; 5, 10, 20, 40, 80, and 160$\sigma$, where 1$\sigma$ is 0.9 mJy beam$^{-1}$) 
		and images of high velocity components of CS (blue and red contours; a 3$\sigma$ interval from 3$\sigma$, where 1$\sigma$ is 2.7 mJy beam$^{-1}$ km s$^{-1}$). 
		(b) Moment 1 (color) and moment 0 maps of CS (contours with a 6$\sigma$ interval from 3$\sigma$, where 1$\sigma$ is 7.5 mJy beam$^{-1}$ km s$^{-1}$). 
		(c) Moment 0 images of SO (color) and CS (black contours). 
		(d) Schematic illustration of the infalling-rotating envelope model. 
		The CS distribution is missing around the region A indicated by a red dashed line due to asymmetry of the gas distribution, 
		while it seems to be concentrated around the region B indicated by a blue dashed line.  
		\label{fig:f1}}
\end{figure}

\begin{figure}
\epsscale{0.7}
\plotone{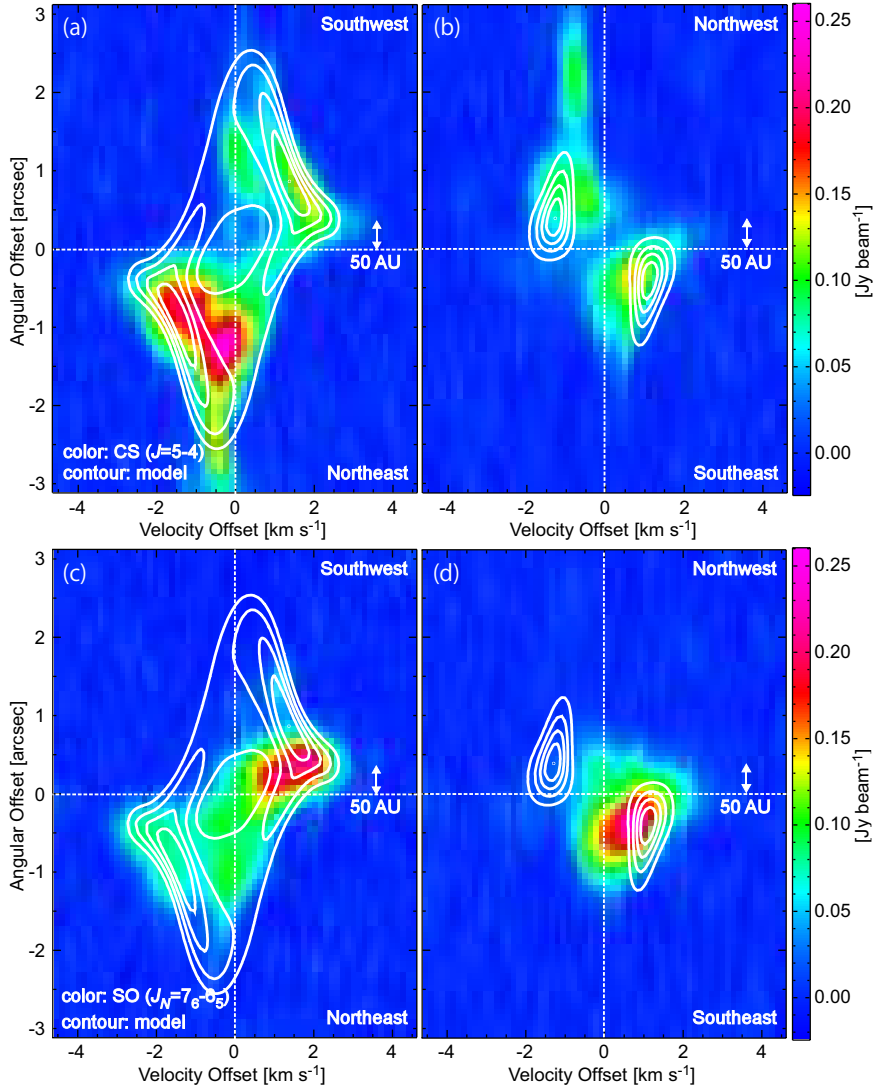}
\caption{(a) Position-Velocity (PV) diagram of CS along the envelope, on which the result of the infalling -rotating envelope model is superimposed (contours). 
		The model contours are every 20 \% of the peak intensity. 
		(b) PV diagram of CS along the line perpendicular to the envelope, on which the result of the model is superimposed (contours). 
		Contours are the same as in (a). 
		(c) PV diagram of SO along the envelope. 
		Contours are the same as in (a). 
		(d) PV diagram of SO along the line perpendicular to the envelope. 
		Contours are the same as in (a). 
		\label{fig:f2}}
\end{figure}


\begin{thebibliography}{}
\bibitem[Aikawa et al. 2012]{aik12} Aikawa, Y., Wakelam, V., Hersant, F., Garrod, R. T., and Herbst, E., 2012, \apj, 760, 40.
\bibitem[Aota, Inoue, and Aikawa 2015]{aot15} Aota, T., Inoue, T., and Aikawa, Y., 2015, \apj, 799, 141.
\bibitem[Aso et al. 2015]{aso15} Aso, Y., Ohashi, N., Saigo, K., 2015, \apj, 812, 27. 
\bibitem[Brinch et al. 2007]{bri07} Brinch, C., Crapsi, A., Jorgensen, J. K., Hogerheijde, M. R., and Hill, T., 2007, A\&A, 475, 915.
\bibitem[Brown and Chandler 1999]{bro99} Brown, D. W. and Chandler, C. J. 1999, MNRS, 303, 855.
\bibitem[Eisner 2012]{eis12} Eisner, J. A. 2012, \apj, 755, 23.
\bibitem[Green et al. 2013]{gre13} Green, J. D. et al. 2013, \apj, 770, 123.
\bibitem[Harsono et al. 2014]{har14} Harsono, D., Jorgensen, J. K., van Dishoeck, E. F., Hogerheijde, M. R., Bruderer, S., Persson, M. V., and Mottram, J. C. 2014, A\&A, 562, A77.
\bibitem[Hogerheijde 2001]{hog01} Hogerheijde, M. R. 2001, \apj, 553, 618.
\bibitem[Ishii and Tamura 2004]{ish04} Ishii, M. and Tamura, M., 2004, \apj, 612, 956.
\bibitem[Li, Krasnopolsky, and Shang 2011]{li11} Li, Z.-H., Krasnopolsky, R., and Shang, H., 2011, \apj, 738, 180.
\bibitem[Lique, F. et al. 2006]{liq06} Lique, F., Dubernet, M. -L., Spielfiedel, A., and Feautrier, N., 2006, A\&A, 450, 399. 
\bibitem[Machida, Inutsula, and Matsumoto 2014]{mac14} Machida, M.N., Inutsuka, S., and Matsumoto, T., 2014, MNRAS 438, 2278.
\bibitem[Murillo et al. 2013]{mur13} Murillo, N. M., Lai, S-P., Bruderer, S., Harsono, D., and van Dishoeck, E. F., 2013, A\&A, 560, A103.
\bibitem[Ohashi et al. 2014]{oha14} Ohashi, N. et al. 2014, \apj, 796, 131.
\bibitem[Ohashi et al. 1997]{oha97} Ohashi, N., Hayashi, M., Ho, P. T. P., Momose, M., Tamura, M., Hirano, N., Sargent, A. A., 1997, \apj, 488, 317.
\bibitem[Oya et al. 2014]{oya14} Oya, Y. et al. 2014, \apj, 795, 152.
\bibitem[Oya et al. 2015]{oya15} Oya, Y. et al. 2015, \apj, 812, 59. 
\bibitem[Saigo and Hanawa 1998]{sai98} Saigo, K. and Hanawa, T. 1998, \apj, 493, 342.
\bibitem[Sakai et al. 2008]{sak08} Sakai, N., Sakai, T., Hitora, T., and Yamamoto, S. 2008, \apj, 672, 371.
\bibitem[Sakai and Yamamoto 2013]{sak13} Sakai, N. and Yamamoto, S. 2013, Chem. Rev., 113, 8981.
\bibitem[Sakai et al. 2014a]{sak14a} Sakai, N., et al. 2014a, Nature, 507, 78.
\bibitem[Sakai et al. 2014b]{sak14b} Sakai, N., et al. 2014b, \apj, 791, L38.
\bibitem[Schoier et al. 2005]{sch05} Sch$\rm \ddot{o}$ier, F. L., van der Tak, F. F. S., van Dishoeck, E. F., and Black, J. H., 2005, A\&A, 432, 369.
\bibitem[Stahler et al. 1994]{sta94} Stahler, S.W., Korycansky, D. G., Brothers, M. J., and Touma, J., 1994, ApJ, 431, 341.
\bibitem[Takakuwa et al. 2012]{tak12} Takakuwa, S., Saito, M., Lim, J., Saigo, K., Sridharan, T. K., and Patel. N. A. 2012, ApJ, 754, 52.
\bibitem[Tobin et al. 2012]{tob12} Tobin, J.~J., Hartmann, L., Chiang, H-F., Wilner, D.~J., Looney, L. W., Loinard, L., Calvet, N.and D$^{\prime}$Alessio, P. 2012, Nature, 492, 83.
\bibitem[Torres et al. 2007]{tor07} Torres, R. M., Loinard, L., Mioduszewski, A. J., and Rodriguez, L. F. 2007, \apj, 671, 1813.
\bibitem[Tsukamoto et al. 2015]{tsu15} Tsukamoto, Y., Iwasaki, K., Okuzumi, S., Machida, M.N., and Inutsuka, S., 2015, ApJ, 810, L26.
\bibitem[Whitney et al. 2003]{whi03} Whitney, B.A., Wood, K., Bjorkman, J.E., and Wolff, M.J., 2003, ApJ, 591, 1049. 
\bibitem[Yen et al. 2013]{yen13} Yen, H.~-W., Takakuwa, S., Ohashi, N., and Ho, P. T. P., 2013, \apj, 772, 22.
\end{thebibliography}
\end{document}